\input{aipcheck}

\documentclass[
    ,final            
  ]
  {aipproc}

\layoutstyle{6x9}

\begin{document}
\def\lya{Ly$\alpha$ }
\def\lbgn{$LBG_N$ }
\def\lbgl{$LBG_L$ }
\title{The physical properties and evolution of Ly$\alpha$ emitting galaxies}


\author{L. Pentericci}
{ address={INAF - Rome Observatory  Via di Frascati 33 Monte Porzio Catone }
}

\author{A. Grazian}
{ address={INAF - Rome Observatory  Via di Frascati 33 Monte Porzio Catone }
}
\author{A. Fontana }
{ address={INAF - Rome Observatory  Via di Frascati 33 Monte Porzio Catone }
}

\begin{abstract}
A significant fraction of high redshift starburst galaxies  presents strong
\lya\ emission. Understanding  the nature of these galaxies is important
to  assess the role they played in the early Universe and to  shed light
on the relation between the narrow band selected \lya\ emitters  and the
Lyman break galaxies: are the \lya emitters a subset of the general LBG
population?  or do  they represent the youngest galaxies in their
early phases of formation?
We studied a sample of UV continuum selected galaxies  from
z$\sim$2.5 to z$\sim$6 (U, B, V and i-dropouts) from the GOODS-South 
survey, that have been observed spectroscopically. Using the 
GOODS-MUSIC catalog we  investigated their physical properties, such as total
masses, ages, SFRs, extinction etc as determined from
a spectrophotometric fit to the multi-wavelength (U band to mid-IR) SEDs, and
their dependence on the emission line characteristics.
In particular we determined the nature of the LBGs with \lya\ in emission and compared them to the properties of narrow band selected \lya\ emitters.
For U and B-dropouts we also compared the properties of LBGs with and without the  \lya\ emission line.
\end{abstract}

\maketitle


\section{The relation between  Ly$\alpha$ emitters and
 Lyman break galaxies}
Over the past few years large samples of   galaxies of
 up to the highest redshifts  have been found (Iye et al. 2006, Kashikawa et al. 2006) using techniques that
 rely on color-selection criteria.
 Among the various methods, one of the more efficient is the Lyman break 
technique (Steidel \& Hamilton 1993), which  is sensitive to the 
presence of the 912 \AA\  break and is effective in finding star-forming galaxies.  This method requires a  blue spectrum, implying low to moderate dust absorption.
 An alternative technique is to search
for \lya\ emission,  through very deep, narrow-band imaging in selected redshift windows,  as first shown  by Cowie \& Hu (1998).
\lya emitters (LAEs) are generally selected to have high restframe \lya\ equivalent width, typically $EW > 20$\AA, with no constraint on the continuum.
 Consequently, this method tends to select much fainter galaxies, compared to the general LBGs population. Many \lya emitters have now been found
(e.g. Ouchi et al. 2004, Gawiser et al. 2007) and  several
distant large-scale structures or protoclusters  have been discovered
(e.g. Venemans et al. 2007).
\\
Each of the two methods suffers from a different selection bias: the two 
resulting populations of galaxies  overlap partially and the  relationship 
 between them is not clear.
Various scenarios have been proposed to explain the properties of Ly$\alpha $ emitters. 
 Because the \lya\ line is easily suppressed by dust, \lya\ emitters are often 
identified as extremely
young galaxies, experiencing their initial phase of  star formation in essentially dust-free environments (e.g.
Gawiser et al. 2007).
However, the different behavior of \lya\ and continuum photons in interacting with dust  makes it possible also for older galaxies to exhibit \lya\ in emission, as predicted e.g. in the models of Haiman \& Spaans (1999). Therefore LAEs (or a fraction of
them) could also represent an older population with active star-forming regions, where the gas kinematics can favor the escape
of \lya\ emission.
This scenario is partially  supported  by the results of Shapley et al. (2003): based on rest-frame optical photometry, 
they  concluded that LBGs with Ly$\alpha $ 
in emission are "old'' (ages greater than a few $\times 10^8$ yr), while "young'' (ages less than $\sim$100 Myrs) LBGs exhibit Ly$\alpha $ in absorption.
In addition  Lai et al. (2007) found that some high-redshift \lya\ galaxies  could be consistent with hosting relatively old stellar population.
Finally, some authors have suggested that galaxies could have more than one
\lya bright emission phase (e.g. Thommes \& Meisenheimer 2005). An initial - primeval - phase in which dust is virtually non-existent, and a later, secondary phase in which strong galactic winds, as observed in some Lyman break galaxies, facilitate the escape of Ly-${\alpha } $ photons after dust has already been formed.
\\
Clearly, it  is worthwhile understanding  the real relation between galaxies with \lya emission and the general LBG population,
 so that properties of the overall high-redshift galaxy population, such as the total stellar mass density, can be betterconst
rained.
\section{The GOODS-MUSIC sample}
To shed light on this issue, we  analyzed a sample of LBGs
selected as U-B-V- and i-dropouts from the GOODS-MUSIC sample (Grazian 
et al. 2006), and with VLT spectroscopic confirmation 
(Vanzella et al. 2006, Vanzella et al. 2008, Popesso et al. 2008). 
For the color cut adopted see  Giavalisco et al. (2004). 
Given the good quality of the spectra, it was possible to assess
whether the galaxies exhibit
\lya emission or the line is  absent and/or appears as an absorption feature.
In the following we define $LBG_L$  the LBGs with line emission
and  $LBG_N$ the LBGs with no line.
For the lower redshift sets (U and B dropouts), the spectroscopic confirmation of LBGs is effective, regardless of the presence or absence of the Ly$\alpha$ line. At higher redshift  ($z >4.8$) 
the spectroscopic confirmations  are almost exclusively
based on the presence of the Ly$\alpha$ line in emission,
while the objects with possible Ly$\alpha$ in absorption (or absent)
become progressively more difficult to identify.
Therefore for V and I-dropouts it is not possible to make a 
comparison of the properties of \lbgn\ and \lbgl, since the first set 
is not complete.
We will therefore compare the properties of \lbgl\ and \lbgn\ for 
galaxies at redshift $\sim$2.5 to $\sim$4.8.
At all redshifts, ($\sim$2.5 to $\sim$6), we will 
study the properties of \lbgl\, 
and  compare them to the properties of the narrow band selected 
 LAEs. 
\\
For all galaxies the physical properties,
 such as total stellar mass and age, star formation rate metalicity and so on,
were derived from a spectrophotometric fit of the multi-wavelength (U band to mid-IR) SEDs, using the templates of Bruzual \& Charlot (2007) (an example is shown in  Figure 1).
A full description of this procedure with a discussion of the uncertainties involved  can be found in Fontana et al. (2006).

\begin{figure}
  \includegraphics[height=.4\textheight]{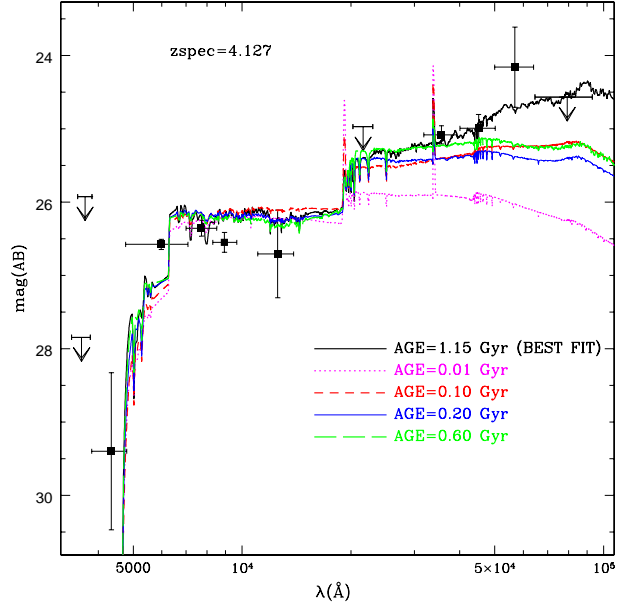}
  \caption{The SED of galaxy with ID8073 at z=4.127: the black line is the best fit with age =1.1 Gyrs. The other lines with
different colors and types represent the bestfit models with varying ages}
\end{figure}

\section{2. Properties of line emitting LBGs at high redshift : 
not just primeval galaxies?}
We first analyzed the properties of all LBGs that show 
the \lya\  line in emission.
In about  half of the galaxies, the rest-frame EW of the line is  
high enough (EW $> 20$ \AA ) 
that they would be selected also in a narrow band survey.
In the rest of the galaxies the EW is weaker than this limit.
Our main results are the following:
\begin{itemize}
\item
Although most galaxies are fit by  young stellar populations,
a small but non negligible fraction
has SEDs that cannot be represented well by young models and
require considerably older stellar component, up to $\sim 1$Gyr.
In Figure 1 we show the SED of one of these bright emission
line galaxies with old age, a galaxy at redshift 4.1:
the best fit model is shown with a black line
(best fit age 1.1 Gyr). The relative best fit models with younger ages
(with age set equal to 10, 100, 200 and 600 Myrs respectively) are also
shown with different colors. They
clearly give a much poorer representation of the observed SED,
expecially in the mid-IR range.
\\
There is no apparent relation between age and EW: some of the oldest galaxies have high line EW, and  should be also selected in narrow-band surveys.
Therefore not all Ly$\alpha$ emitting galaxies
are primeval galaxies in the very early stages
of formation, as is commonly assumed.
\item
We  find a range of stellar populations, with masses from $5\times 10^8 M_\odot$ to $5 \times 10^{10} M_\odot$  and SFR from few to  60$M_\odot yr^{-1}$.
These values are higher than those derived in general for narrow band selected LAEs.
Clearly most (but not all) of the NB emitters have somewhat
fainter continuum  than our LBGs: we are studying objects that are, on average,
 intrinsically brighter and thus more massive.
However the difference in mass is larger than expected:
our sample comprises also
galaxies with similarly faint broad-band magnitudes,
thanks to the very deep  GOODS observations.
The difference  could be due, in part, to a trend that we observe in 
Figure 2. Here we show how the stellar mass and
\lya\ EWs  are related. The dashed lines indicate the median mass of each sample. Although there is no net correlation between mass and EW,
we find a significant  lack of massive galaxies with high EW, which could be explained if the most massive galaxies
were either dustier and/or if they contained more neutral gas than less massive objects.
\item
Finally we find that more than half of the galaxies contain small but 
non negligible amounts of dust:
the mean  E(B-V) derived from the SED fit and the  EW are
well-correlated, although with a large scatter, as already found at lower redshift.
 \end{itemize}
The results presented here are extensively discussed in Pentericci et al. 2008

\begin{figure}
  \includegraphics[height=.25\textheight]{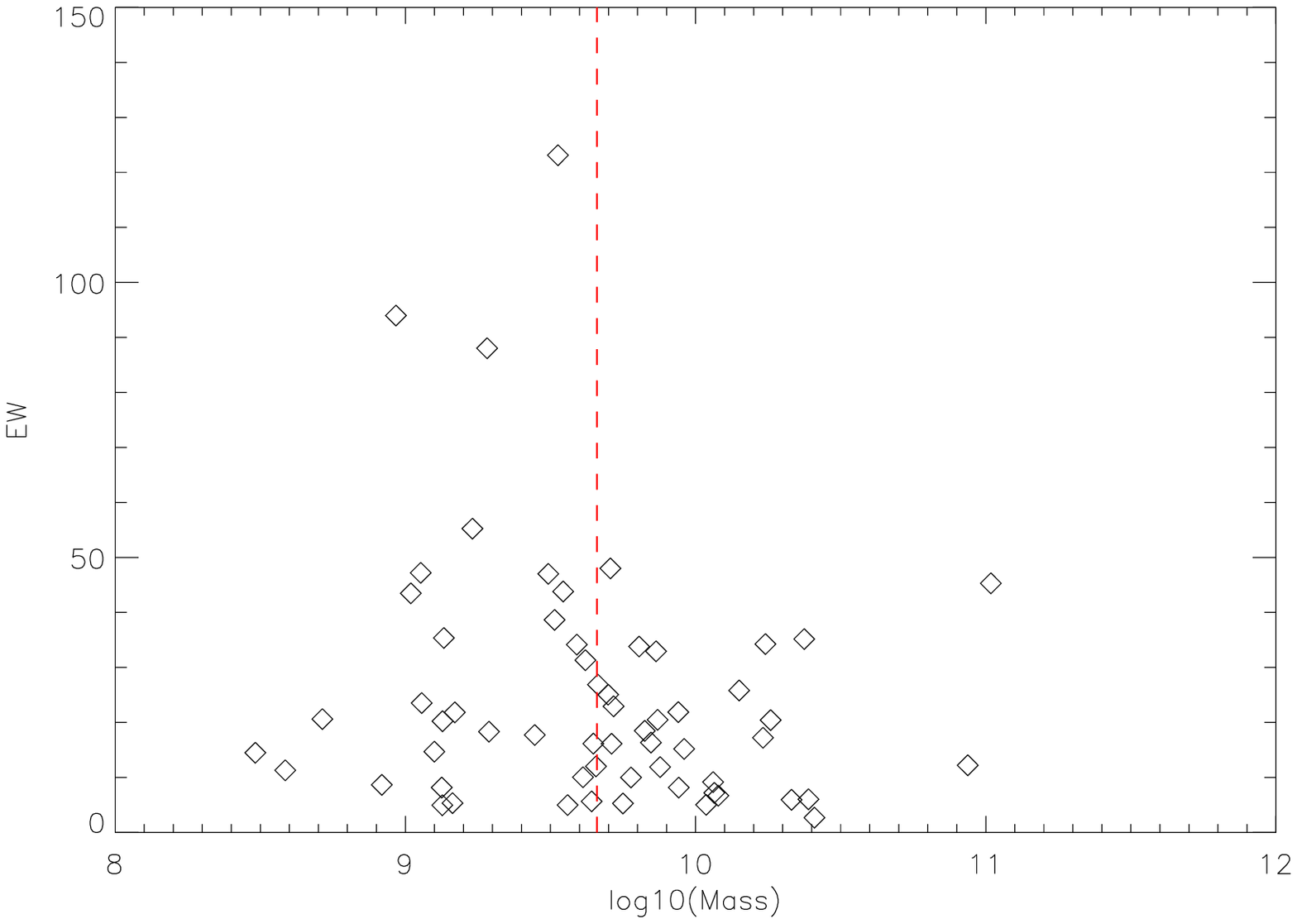}
  \includegraphics[height=.25\textheight]{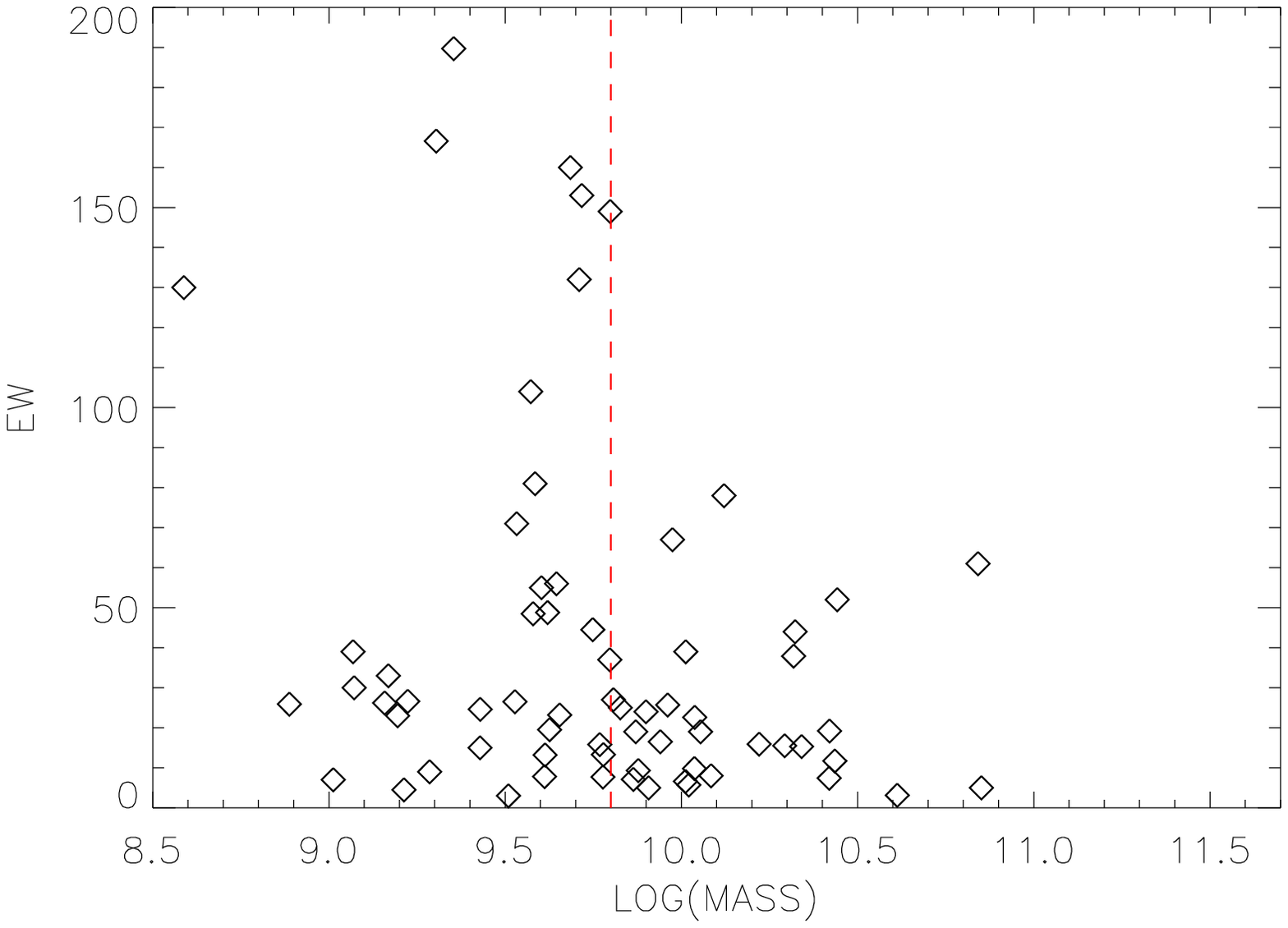}
  \caption{The dependence of stellar masses  (best fit values)
on  \lya\ EW. The left figure refers to the U-dropout sample , while the right figure contains B-V- and i-dropouts. The dashed lines indicate in each case
 the median masses. }
\end{figure}

\begin{figure}
  \includegraphics[height=.5\textheight,clip]{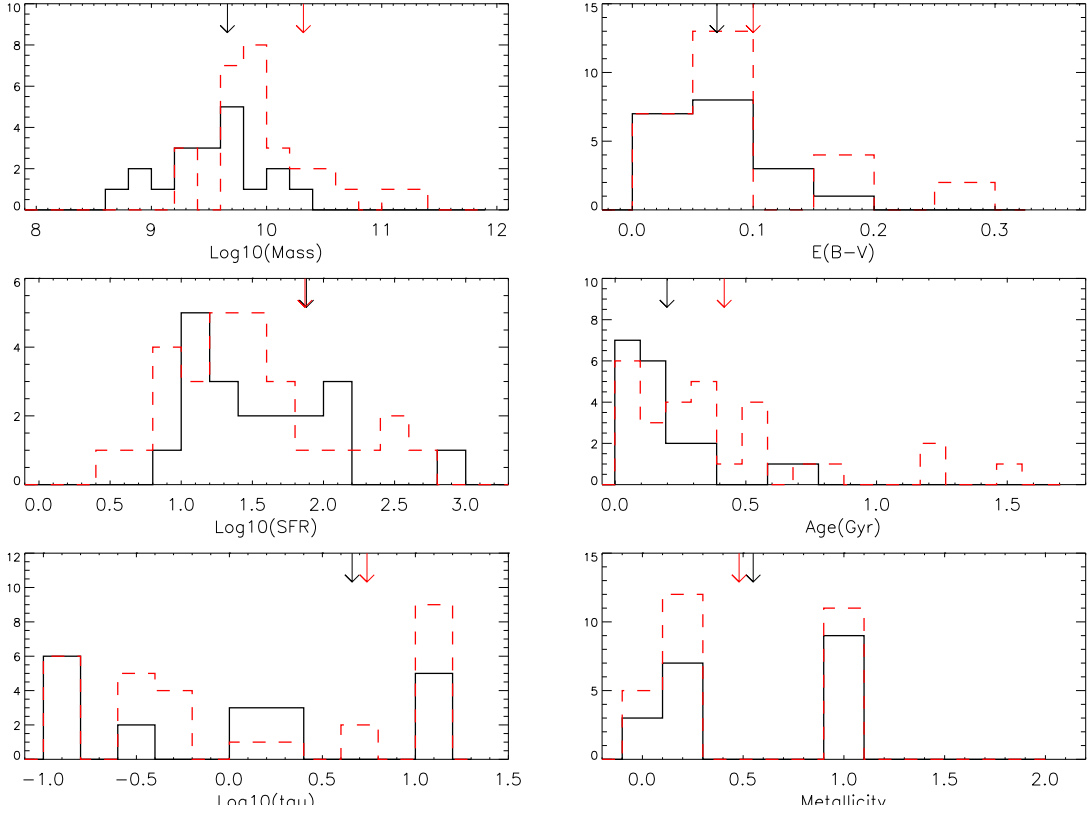}
  \caption{The physical properties of B  and V-dropouts. Top left: distribution of total stellar mass; top right the extinction parameter
E(B-V) derived from the spectral fitting; middle left the total star formation rate derived from the spectral fitting; middle ri
ght: stellar ages; bottom left: $\tau$ the star formation e-folding time-scale; bottom right: the metalicity. Solid black and dashed light histograms denote \lbgl\ and \lbgn\, respectively. The blck and light arrows indicate the mean values of each sample
.}
\end{figure}
\section{2. Comparison between LBGs with and without emission line}
We then compared the properties of \lbgl\ to those of \lbgn.
As already detailed in the introduction, this is only possible (in an unbiased way) for the lower redshift part of the sample.
In particular our analysis  of the B-dropouts was presented in 
Pentericci et al. (2007), while the analog study for 
the U-dropouts will be reported in Pentericci et al. 2009 (in preparation).
\\
Our main results are the following:
\begin{itemize}
\item
Both the total stellar masses  and the median ages are  considerably
lower for the \lbgl\ compared to the \lbgn.
For B dropouts the average mass is $(5.0 \pm 1)\times 10^9 M_{\odot}$ for the \lbgl\ and
$(2.3 \pm 0.8) \times 10^{10}  M_{\odot}$
for the others, i.e. a factor of almost five  higher.
The K-S test gives a very low value, implying that the two
populations are different from each other
with $>99.8 \%$  probability. We can therefore conclude that the
\lbgl\ are less massive than the \lbgn\ by a factor of almost 5.
The median ages are also quite different, with an average  of $200\pm 50$ Myr
for the \lbgl,  an age  distribution that  is very peaked
towards low age values  and is  basically confined to values  below 300 Myr.
The \lbgn\ on the other hand have an average age of $410\pm 70$ Myr
(older by a factor of more than 2)
and there are several galaxies with ages exceeding 1 Gyr, which is a considerable fraction of the cosmic time at redshift $\sim 4$.
Again performing a K-S test,  the two populations are
different with a  probability $>98 \%$ .
We conclude that the \lbgl\ are significantly younger galaxies 
than the \lbgn.
\item
As at redshift 4, also at redshift 3 for U dropouts
the masses of $LBG_N$ are larger than the masses of $LBG_L$ and the galaxies
without emission line are older.
The average values 
differ considerably (the same holds true if one derives
the median value instead of the average).
The  K-S test gives a significance of   more than 99$\%$ in
both cases (P=0.0003 for age and P=0.000 for mass).
However we notice that the difference between the samples is less
 pronounced than
at higher redshift: the mass difference is
less  than a factor of 2, and the age difference is only a 50\% factor.
Expecially for the age, we find that although the two distributions are different,
the range spanned by \lbgl and \lbgn is basically equal.
In other words LBGs seem to be a more uniform population at lower redshift and
the properties vary  less with the presence or absence of \lya\ emission.
\item
\lbgl\ are less dusty than \lbgn\, although the difference is not large. 
This can be also observed  from the UV slope of the stacked  spectra that
we derived for the z$\sim 3$ \lbgl\ and \lbgn\ separately (Pentericci et al. 2009).  
In general, as a natural consequence of the initial color selection, 
all galaxies contain small amounts of  dust.
\item  Both populations are forming stars very actively and the average star formation rates similar.
However for U-dropouts  we find that, while in  \lbgl\ the current SFR is approximately equal to the past average SFR (derived as the total assembled mass divided by the galaxy age), for \lbgn\ this is not true. For many \lbgn\ the past average SFR is much larger than the current value, indicating  that in the past they have formed stars much more vigorously than at present.
\item
The morphological properties are also somewhat different: 
\lbgl\ tend to be more nucleated than those without line emission.
Most of the absorbers have a very diffuse and/or clumpy morphology: this 
could be due either to the presence of more dust or to 
intrinsic morphological differences.
\end{itemize}

\begin{figure}
  \includegraphics[height=.3\textheight,clip]{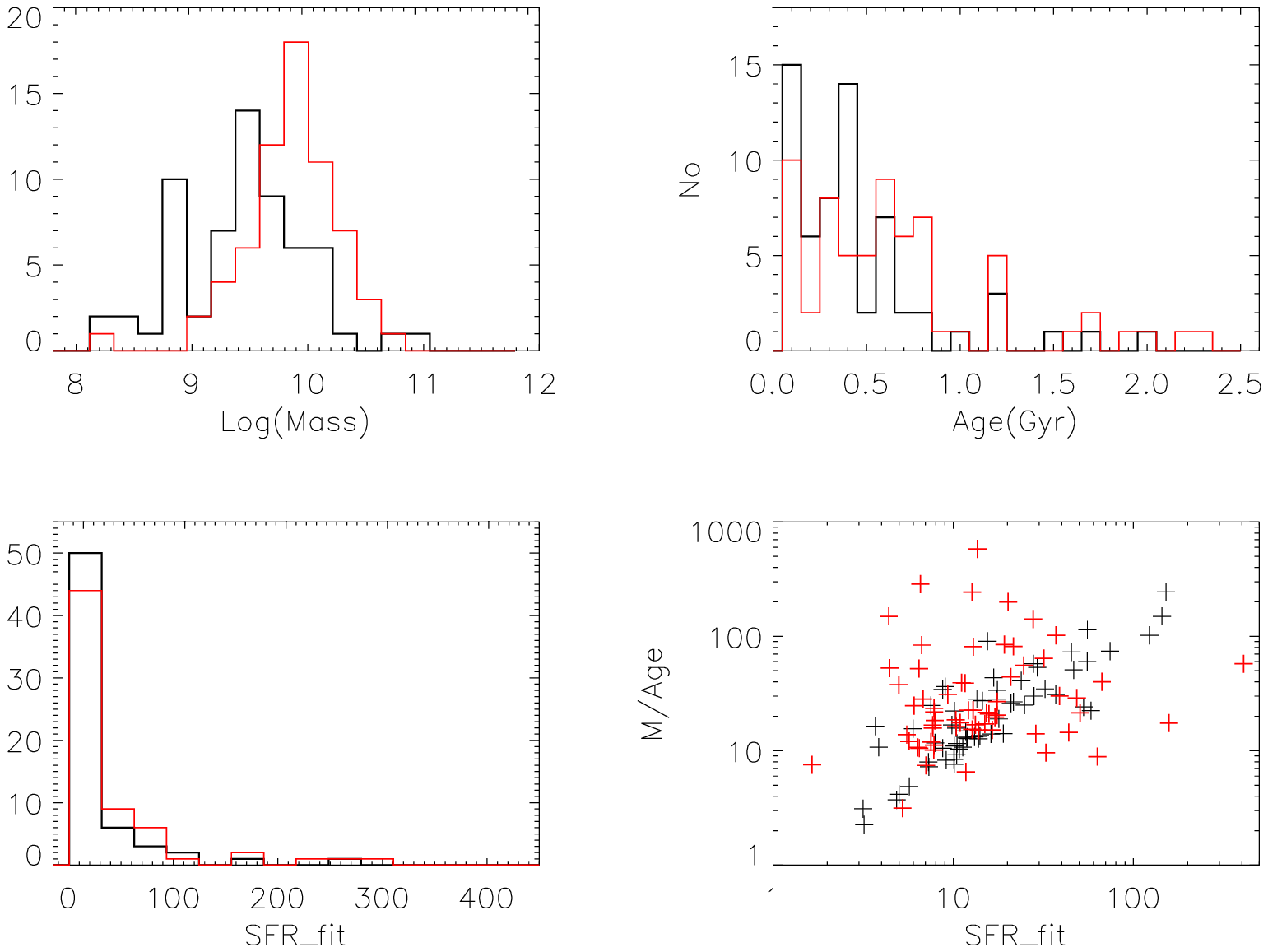}
  \caption{Same as in figure 3 for U-dropouts and for mass and age only}
\end{figure}


\begin{thebibliography}{18}
\bibitem{cowie} Cowie, L.~L., \& Hu, E.~M.\ 1998, AJ, 115, 1319
\bibitem{F06} Fontana, A., Salimbeni, S., Grazian, A., et al. 2006 A\&A 
\bibitem{gaw07} Gawiser E., et al.\  2007, ApJ, 671, 278
\bibitem{gia04} Giavalisco, M., et al\ 2004, ApJL, 600, L103
\bibitem{grazian06a} Grazian, A., et al. 2006  A\&A  449, 951
\bibitem{hai99} Haiman, Z., \& Spaans, M.\ 1999, ApJ, 518, 138
\bibitem{iye06} Iye, M., et al.\ 2006,Nature, 443, 186
\bibitem{kas06} Kashikawa, N., et al.\ 2006, ApJ, 648, 7
\bibitem{lai} Lai, K., et al.\ 2007, ArXiv e-prints, 710, arXiv:0710.3384
\bibitem{ou04} Ouchi, M., et al.\ 2004, ApJ, 611, 660

\bibitem{pen07} Pentericci, L. et al.\ 2007, A\&A, 471, 433
\bibitem{pen08} Pentericci, L. et al.\ 2008,  A\&A, arXiv0811.1861P
\bibitem{pop08} Popesso, P., et al.\ 2008, ArXiv e-prints, 802, arXiv:0802.2930
\bibitem{stei93} Steidel, C.~C., \& Hamilton, D.\ 1993, AJ, 105, 2017
\bibitem{tom05} Thommes, E., \& Meisenheimer, K. 2005, A\&A, 430, 877
\bibitem{van08} Vanzella, E., et al.\ 2008,  A\&A, 478, 83
\bibitem{van06} Vanzella, E., et al.\ 2006, A\&A, 454, 423
\bibitem{ven07} Venemans, B.~P., et al.\ 2007,  A\&A, 461, 823
\end{thebibliography}
\end{document}